\documentclass[conference]{IEEEtran}
\usepackage{graphicx}
\usepackage{booktabs}
\usepackage{multirow}
\usepackage{amsmath,amssymb}
\usepackage{url}
\usepackage{xcolor}
\usepackage{algorithm}
\usepackage{algorithmic}
\usepackage{array}
\usepackage{cite}
\usepackage{graphicx, tikz}
\usetikzlibrary{arrows.meta} 

\usepackage{float}
\usepackage{needspace}
\usepackage{subcaption}
\newcommand{\AIaaS}{{AIaaS}}
\newcommand{\AIPaging}{{AI-Paging}}
\newcommand{\QoS}{{QoS}}

\usepackage{tikz}
\usetikzlibrary{arrows.meta,positioning,fit,calc}
\usepackage{subcaption}

\newcommand{\AISI}{{AISI}}     
\newcommand{\AIST}{{AIST}}     
\newcommand{\ASP}{{ASP}}       
\newcommand{\COMMIT}{{COMMIT}} 
\newcommand{\EVI}{{EVI}}       

\newcommand{\AEXF}{{AEXF}}     

\title{AI-Paging: Lease-Based Execution Anchoring for Network-Exposed AI-as-a-Service}

\author{
    Mohaned Chraiti \IEEEauthorrefmark{1}, and Merve Saimler \IEEEauthorrefmark{2}\\
    \IEEEauthorrefmark{1}Electronics Engineering Department, Sabancı University, Türkiye\\
    \IEEEauthorrefmark{2} Ericsson Research, Türkiye\\
    
    Emails: mohaned.chraiti@sabanciuniv.edu and merve.saimler@ericsson.com.
}

\begin{document}
\maketitle

\begin{abstract}


With AI-as-a-Service (AIaaS) now deployed across multiple providers and model tiers, selecting the appropriate model instance at run time is increasingly outside the end user’s knowledge and operational control. Accordingly, the6G service providers are envisioned to play a crucial role in exposing AIaaS in a setting where users submit only an intent while the network helps in the intent-to-model matching (resolution) and execution placement under policy, trust, and Quality of Service (QoS) constraints. The network role becomes to discover candidate execution endpoints and selects a suitable model/anchor under policy and QoS constraints in a process referred here to as \emph{AI-paging} (by analogy to cellular call paging). In the proposed architecture, AI-paging is a control-plane transaction that resolves an intent into an AI service identity (AISI), a scoped session token (AIST), and an expiring admission lease (\COMMIT) that authorizes user-plane steering to a selected AI execution anchor (AEXF) under a QoS binding.AI-Paging enforces two invariants: \textbf{(i) lease-gated steering} (without \COMMIT, no steering state is installed) and \textbf{(ii) make-before-break anchoring} to support continuity and reliability of AIaaS services under dynamic network conditions. We prototype AI-Paging using existing control- and user-plane mechanisms (service-based control, \QoS\ flows, and policy-based steering) with no new packet headers, ensuring compatibility with existing 3GPP-based exposure and management architectures, and evaluate transaction latency, relocation interruption, enforcement correctness under lease expiry, and audit-evidence overhead under mobility and failures.

\end{abstract}


\section{Introduction}





Communication Networks are being re-shaped by two converging shifts: (i) service providers are exposing connectivity through standardized {network application-facing interfaces (APIs)} (rather than bespoke integrations), and (ii) Artificial Intelligence (AI) capabilities are increasingly embedded into the network and exposed as consumable services (AI-native 6G networks). The first shift manifests through 3GPP exposure functions (NEF), APIs on-boarding/discovery (CAPIF), service layer enablers (SEAL/AIMLE), and ecosystem APIs (CAMARA), together covering network services exposure, network data exposure, QoS exposure, and MLOps-oriented capabilities~\cite{gsma_open_gateway,camara_project,3gpp_capif_ts23222}. The second shift is more structural: AI services are not merely “applications running at the edge". They need infrastructure that supports model management (selection and placement), adaptive response to network load, and enforcement of service semantics, with lifecycle-managed AI built on MLOps-as-a-Service and network-integrated intelligence. This turns AI-as-a-Service (\AIaaS) into a control and enforcement problem rather than a hosting problem  \cite{Merve1, merve2, SaimlerSlides2024}. In such a set up, a network-exposed \AIaaS\ request is naturally expressed as an intent, requiring the network to provide an endpoint-agnostic service handle with QoS-enforced execution, continuity under mobility/overload/failures, and audit-ready evidence for cross-domain compliance and accountability.\footnote{The compliance re-enforces data governance and provenance requirements, supporting emerging regulatory and privacy frameworks (e.g., EU AI-Act \cite{eu_ai_act_2024}–like governance models).}



The gap is a missing \emph{protocol primitive} that connects application intent to enforceable user-plane behavior. Current ML serving systems are optimized around inference throughput and tail latency inside a compute substrate (e.g., batching/caching, pipeline scaling, model selection)~\cite{clipper_nsdi,inferline_osdi}. They do not define a network-level contract for (a) when steering is authorized, (b) how steering state is revoked on expiry, or (c) how execution can relocate without breaking the client-visible binding. Conversely, network exposure frameworks define how applications call service provider APIs, but do not define execution anchoring, admission leases, or make-before-break relocation as first-class service semantics~\cite{3gpp_capif_ts23222,gsma_open_gateway}.
As a result, AIaaS today deployments either bind clients to endpoints (fragile under mobility/load) or rely on best-effort steering (silent SLO violations), with limited auditability.


Within the initiatives of developing a control-plane transaction that operationalizes AIaaS orchestration within AI-native networks ( \AIaaS\ enforceable as a network service), this paper introduces {\AIPaging}, a mechanism conceptually analogous to mobility paging. While the latter resolves {where a user equipment can be reached}, AI-Paging resolves {where an intent can be executed} while preserving stable service identity. Given an intent and service constraints, AI-Paging produces three artifacts with explicit semantics: (i) a AI service identity (\AISI), (ii) a scoped session token coped session token (\AIST), and (iii) an expiring admission lease (\COMMIT) that authorizes user-plane steering to a selected AI execution anchor (\AEXF) under a specified \QoS\ binding. This separation enables lifecycle-managed, QoS-assured AI execution without exposing internal network or AI infrastructure details to applications. The design is anchored by two safety invariants: {(i) lease-gated steering}—no valid \COMMIT\ implies steering state must not exist—and {(ii) make-before-break anchoring}—relocation installs a new lease and steering state before draining and releasing the old anchor.

The contributions of this paper are summarized as follows.
\begin{itemize}
  \item We propose AI-Paging, a lease-based execution-anchoring transaction for network-exposed \AIaaS, with explicit service artifacts and enforceable invariants.
  \item We develop a minimal artifact model (\AISI/\AIST/\COMMIT/\EVI) and an end-to-end procedure (bind--page--commit--steer--serve--move) that defines a verifiable control-to-data-plane contract.
  \item We build a prototype that maps the artifacts to existing 5G control- and user-plane mechanisms (service-based control interfaces, \QoS\ flows, policy-based steering) without new packet formats~\cite{3gpp_5gs_arch_23501,3gpp_5gs_proc_23502}.
  \item We evaluate latency/success, relocation continuity, lease-expiry enforcement correctness, robustness under overload/failure, and evidence overhead.
\end{itemize}

\section{Background and Related Work}

Network-exposed \AIaaS\, as envisioned in AI-native 6G networks, sits at the intersection of two complementary yet individually insufficient lines of work: (i) {standardized network capability exposure and service enablement frameworks}, and (ii) {AI execution and serving}, along with {edge compute platforms}, that enable efficient, scalable training and inference.


\subsection{Exposure and orchestration substrates}
3GPP CAPIF and similar exposure functions standardize how applications onboard, discover, and invoke service provider APIs~\cite{3gpp_capif_ts23222,etsi_ts123222}.More broadly, Open Gateway and CAMARA extend similar principles into a multi-operator, developer-facing ecosystem for exposing network capabilities via harmonized APIs.~\cite{gsma_open_gateway,camara_project}. These frameworks primarily address secure {API exposure}, {authorization}, and interoperability for network services; they do not yet define how AIaaS execution is anchored, admitted with explicit service guarantees, or maintained under network dynamic and failures as part of a standardized \AIaaS\ contract. From a 3GPP SA6 service-layer perspective, 3GPP SEAL and AIML Enablement (AIMLE) define service-layer AI enablement and client participation. AI-Paging complements these capabilities by introducing enforceable admission lease and relocation semantics that bind AI service invocation to network-enforced continuity guarantees.


\subsection{Inference serving systems, steering primitives, and closed-loop automation}
Prediction serving systems such as Clipper and InferLine address the compute-side problem of meeting tail-latency targets through batching/caching, pipeline planning, and autoscaling ~\cite{clipper_nsdi,inferline_osdi}. Similarly, 3GPP NWDAF provides analytics and predictive insights that can supply feasibility predictors (e.g., latency and load prediction) used for candidate anchor ranking~\cite{3gpp_nwdaf_ts23288}. More broadly, closed-loop automation and analytics (e.g., zero-touch management architectures) enable telemetry-driven optimization and policy feedback loops~\cite{etsi_zsm_002}. These approaches enhances and safeguards the efficiency of AI/ML execution once computational placement decisions and service endpoints have been determined. However, it operates beneath the network exposure layer and therefore does not specify mechanisms for the admission, governance, or relocation of AIaaS services when these are offered as network-exposed services via NEF, SEAL, or CAMARA. In particular, they fall short of defining a transactional primitive that binds an intent to enforceable data-plane behavior with time-bounded admission and make-before-break relocation. AI-Paging complements these components by introducing an explicit admission lease as the sole authority for \AIaaS\ enforcement state, and by defining the relocation semantics required for carrier-grade continuity.

\section{AI-Paging Design: Lease-Based Execution Anchoring}

In this section, we introduce \AIPaging\ as a network primitive to make AIaaS enforceable as a service tier. The design is intentionally conservative: it introduces a small set of artifacts with explicit semantics and ties them to a control-to-data-plane contract that can be implemented using existing steering and \QoS\ primitives. Optimization policies (which model tier is best, which anchor is cheapest) are explicitly not baked into the primitive; they can evolve without changing the service semantics. What is fixed is the enforcement boundary: when the network is authorized to install user-plane state, when it must revoke it, and how it preserves continuity when execution anchors change.

\subsection{Problem and Design Requirements}

Network-exposed AIaaS is conceptually simple but difficult to operationalize as a carrier-grade service. Ideally, an application specifies an intent—an outcome plus constraints such as latency and reliability targets, locality and trust requirements (security, privacy, data governance, execution provenance), and a budget—and the network resolves this intent by selecting an appropriate model tier, deciding where it runs, and provisioning delivery so the agreed tier is enforced. In this context, the model tier determines which AI model variant is selected (e.g., lightweight versus high-capacity models, or cost-versus-accuracy trade-offs), while the execution anchor determines where the selected model tier runs, such as on the device, at the edge, or in centralized cloud resources. Trust and locality requirements can be enforced in practice through mechanisms such as restricting data processing to authorized geographic regions, executing workloads only on certified infrastructure, using attestation mechanisms to verify runtime integrity, and applying governance policies that constrain which datasets or model versions may be used.
Provisioning delivery further requires binding the admitted service instance to deterministic network behavior that satisfies its service-level objectives. In telecom systems, this QoS binding can be realized through QoS flow handling (e.g., 5QI-based QoS flows), network slicing constructs, and policy control mechanisms, ensuring that traffic associated with the AI service is consistently steered and treated according to its latency, reliability, and throughput requirements.


\subsubsection{Why endpoint-bound serving fails as a network service}
A carrier-grade service cannot expose infrastructure churn to applications. Binding a client to a concrete endpoint makes changes in serving location visible whenever mobility, load, maintenance, or failures trigger rebalancing, forcing session re-establishment, context rebuilding, or silent quality degradation. This is not a pathological scenario; it is the expected operating regime of wide-area mobile systems. A network service must therefore expose a stable service identity that persists while the execution anchor changes, preserving the client-visible binding in the same way mobility preserves identity while attachment points evolve.



\begin{figure}[t]
  \centering
    \includegraphics[width=.97\linewidth]{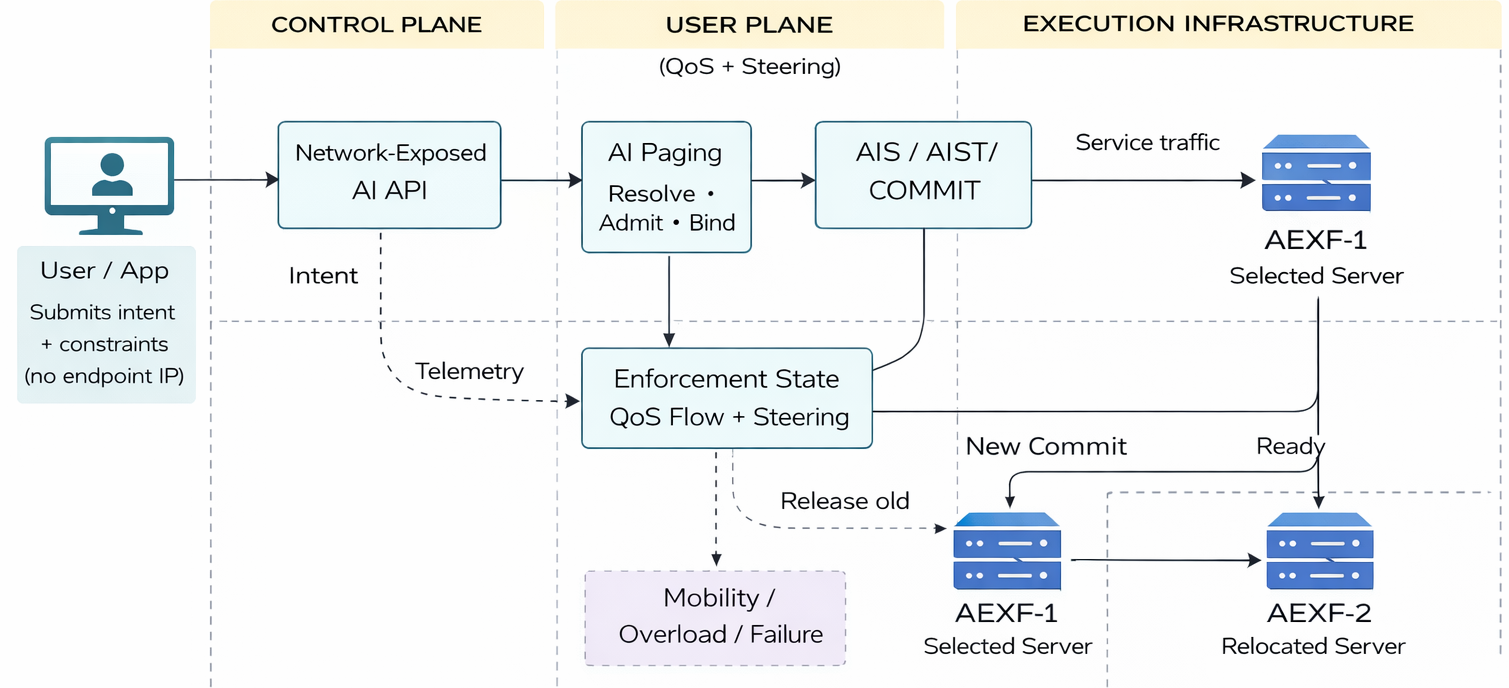} 
  \caption{AI Paging with network-exposed service binding.}
  \label{fig:aipaging_overview}
\end{figure}

\subsubsection{Enforceability and continuity require explicit admission and transactional relocation}
Even with steering mechanisms in place, ``route to an anchor'' is not equivalent to an enforceable service tier.
Service delivery can be dominated by compute queuing, time-varying transport congestion, and policy constraints that exclude otherwise attractive placements.
Without an explicit admission artifact that authorizes steering and expires, there is no sharp boundary between an admitted, enforceable tier and best-effort operation; the system can keep routing traffic while silently violating the contracted tier, and it lacks a deterministic mechanism to revoke enforcement state when the conditions that justified the decision no longer hold.

Continuity under relocation introduces a second enforceability constraint: switching anchors must be transactional.
A steering flip that occurs before the new anchor is admitted and ready risks blackholing traffic; releasing the old anchor too early risks unnecessary interruption.
Robust systems therefore require make-before-break semantics: the new path is admitted and installed before the old path is drained and released, with a bounded overlap window that limits transient inconsistency.
When relocation is implemented through retries and timeouts, continuity becomes an emergent property rather than a guaranteed one.

Fig.~\ref{fig:aipaging_overview} summarizes these gaps by contrasting endpoint-bound serving with a lease-based service primitive that separates stable service identity from the serving anchor, gates steering on explicit admission, and relocates execution using make-before-break semantics. This motivates the design objective pursued in the rest of the paper: introduce the smallest network primitive that turns intent resolution into an admission-backed transaction with artifacts that are enforceable in the user plane and auditable across domains.
The primitive introduced next, \AIPaging, produces a stable service identity and a time-bounded admission lease (\COMMIT) that is the sole authority to install user-plane steering and \QoS\ binding toward an execution anchor, while preserving continuity through transactional relocation and generating evidence records that bind observed delivery to the active lease and anchor.



\subsection{Artifacts and service semantics}
\AIPaging\ resolves an intent into artifacts that separate identity, authorization, admission, and accountability.
These artifacts are the only interface the design assumes between the application-facing control plane and the enforcement mechanisms in the user plane.

\begin{table}[t]
\centering
\vspace*{0.1cm}
\caption{Core \AIPaging\ artifacts and their operational semantics.}
\label{tab:aipaging_artifacts}
\begin{tabular}{@{}p{0.20\linewidth}p{0.72\linewidth}@{}}
\toprule
Artifact & Semantics (what it \emph{means} operationally) \\
\midrule
\AISI\ & Stable service identifier; persists across anchor changes and is the client-visible handle. \\
\AIST\ & Scoped authorization bound to \AISI\ and policy constraints (e.g., allowed tier/locality, expiry). \\
\ASP & Enforceable service contract derived from intent and operator policy” is good but still imo you should explicitly list typical ASP fields: target latency, max jitter/loss, locality region, allowed fallback tier(s), evidence requirements, max relocation rate, lease duration. \\
\COMMIT\ & Time-bounded admission lease; the sole authority to install and maintain steering/\QoS\ state toward a specific anchor \AEXF. \\
\EVI & Evidence record (or pointer) binding observed delivery to \AISI\ and the active \COMMIT\ (and thus to the serving anchor and tier). \\
\bottomrule
\end{tabular}
\end{table}

\AISI\ provides the stable binding that endpoint-based serving lacks: applications bind to a service identity rather than a concrete endpoint, while the network is free to change the serving anchor behind that identity.
\AIST\ scopes authorization to a service session and carries constraint context that prevents the exposure interface from becoming a generic ``open proxy'' to inference infrastructure.
\ASP\ is the enforceable contract derived from the intent; it encodes what the network is willing and able to guarantee in operational terms.
The pivotal artifact is \COMMIT\: it is an expiring admission lease that authorizes steering to a particular execution anchor \AEXF under a specific \QoS\ binding. Finally, \EVI\ binds observed delivery to the service identity and the active lease, enabling dispute-ready accountability without requiring disclosure of internal topology or proprietary scheduling logic.

Fig.~\ref{fig:aipaging_overview} illustrates how these artifacts connect intent resolution to user-plane enforcement. The artifacts are defined in Table I. 

\subsection{Control-to-data-plane contract}
The design is anchored by two correctness conditions that eliminate best-effort ambiguity and make the system testable. A first condition is lease-gated enforcement.
User-plane steering and \QoS\ treatment for a given AIaaS tier are permitted to exist only when backed by a valid, unexpired COMMIT. Operationally, this means that enforcement state is installed with a lease identifier and expiry, and is removed deterministically when the lease expires or is revoked. This creates a sharp boundary between an admitted, enforceable service and best-effort operation.

A second condition is transactional relocation.
Execution anchors may change due to mobility, load, failures, or maintenance, but relocation must be make-before-break: a new lease and enforcement state for the target anchor must exist before the old anchor is drained and released.
The overlap window is explicitly bounded by a drain timeout, so continuity is a correctness property rather than an emergent consequence of retries.

These conditions are orthogonal to any candidate ranking policy.
They define when enforcement is authorized and how enforcement changes without service breakage.

\subsection{AI-Paging transaction}
AI-Paging is realized as a single control-plane transaction whose outcome is either (i) an enforceable service instance, or (ii) a rejection with an actionable cause.
The transaction begins when the client submits an intent.
The control plane derives an enforceable \ASP under policy and issues \AISI\ and \AIST, making the service identifiable and authorized without installing any user-plane enforcement state.
Candidate anchors and eligible model tiers are then selected by feasibility under \ASP constraints; hard constraints (locality/provenance and policy eligibility) filter candidates, while feasibility predictors provided by NWDAF and network telemetry (transport conditions, anchor load/health) rank them.
Admission is finalized by acquiring a \COMMIT\ for the chosen candidate.
Only after a valid lease exists does the system install user-plane classification and steering, mapping traffic associated with \AISI/\AIST\ to the admitted anchor \AEXF under the \QoS\ binding carried by the lease.
During operation, \EVI records are emitted and bound to the active lease, capturing delivery observables needed for compliance verification and attribution.


\subsection{Relocation, timers, and evidence binding}
Relocation is triggered when the current anchor becomes suboptimal or infeasible due to mobility-induced path changes, overload, degraded health, or failures.
The relocation logic preserves the client-visible handle by keeping \AISI\ stable and changing the serving anchor behind it.
The procedure re-runs candidate selection under the existing ASP (including any permitted tier downshift), obtains a new \COMMIT\ for the target anchor, installs new enforcement state, performs an atomic priority switch, drains the old path for a bounded overlap window, and then releases the old lease.
The bounded overlap is controlled by a drain timer, which limits transient duplication and reordering while providing an operational mechanism to balance service continuity against resource overhead. Client-side handling is required but is out of scope.

\Needspace{1\baselineskip}   
\vspace*{0.001pt}   
\begin{algorithm}[H]
\caption{AI-Paging transaction (enforceable intent-to-execution)}
\label{alg:aipaging_txn}
\begin{algorithmic}[1]
\STATE \textbf{Input:} intent $\mathsf{I}$, service provider policy $\Pi$, certified candidates, commit timeout $T_C$
\STATE Derive enforceable \ASP under $\Pi$; issue \AISI\ and \AIST
\STATE Generate and rank feasible candidates (model tier, anchor) under \ASP constraints
\STATE Initialize cause statistics $\mathcal{C}\leftarrow\emptyset$
\WHILE{time $< T_C$ and candidates remain}
  \STATE Select next-best candidate
  \STATE Request admission lease \COMMIT\ for the candidate; receive accept/reject with cause
  \IF{accepted}
    \STATE Install steering/\QoS state bound to \COMMIT; enter serving; start emitting \EVI
    \STATE \textbf{return} SUCCESS$(\AISI,\AIST,\COMMIT)$
  \ELSE
    \STATE Update $\mathcal{C}$ with reject cause; optionally adjust candidate set using allowed fallback in \ASP
  \ENDIF
\ENDWHILE
\STATE \textbf{return} REJECT$(\AISI,\mathcal{C})$
\end{algorithmic}
\end{algorithm}
\vspace{0.5cm}
\begin{algorithm}[h!]
\caption{Make-before-break relocation (transactional anchor move)}
\label{alg:aipaging_move}
\begin{algorithmic}[1]
\STATE \textbf{Input:} \AISI, current anchor $a_0$, current \ASP, drain timeout $T_D$
\STATE Select feasible target anchor $a_1$ under \ASP (and permitted fallback)
\STATE Obtain new admission lease \COMMIT$_1$ authorizing $a_1$ (Algorithm~\ref{alg:aipaging_txn} restricted to relocation)
\STATE Install steering/\QoS state for $a_1$ bound to \COMMIT$_1$
\STATE Atomically flip steering priority to $a_1$
\STATE Drain old path for $T_D$; release old lease and enforcement state for $a_0$
\STATE Emit \EVI event linking the relocation to (\AISI, \COMMIT$_0$, \COMMIT$_1$)
\end{algorithmic}
\end{algorithm}
\vspace{0.4cm}

Evidence is treated as a first-class output since \AIaaS\ is increasingly multi-domain and dispute-prone.
\EVI\ records bind observed delivery to \AISI\ and the active lease identifier, allowing concrete questions to be answered post hoc: which lease authorized steering at the time of a violation, which anchor served the requests, and whether a relocation coincided with transient degradation.
The design does not require disclosure of internal topology or proprietary schedulers; it requires attributable evidence tied to service identity and lease state.

Finally, the design does not mandate a new packet header.
It requires a stable classifier that can be matched by the data plane and mapped to \AISI/\AIST\ and hence to an active \COMMIT.
In practice, this can be realized using an application token, a transport-layer identifier, or an service provider-defined encapsulation, as long as it enables deterministic mapping from user-plane traffic to lease-backed enforcement state.
This keeps \AIPaging\ implementable atop service-based control interfaces and user-plane steering/\QoS primitives in modern mobile cores~\cite{3gpp_5gs_arch_23501,3gpp_5gs_proc_23502}.

\section{Prototype and Protocol-Stack Mapping}

The goal of the prototype is not to introduce new packet formats or a new ``AI layer,'' but to demonstrate that \AIPaging\ can be realized by composing mechanisms that already exist in modern service provider stacks: service-based exposure for control-plane transactions, enforceable \QoS\ constructs, programmable user-plane steering, and telemetry pipelines.
Accordingly, the prototype is organized around a single question: given the artifacts produced by \AIPaging\ (\AISI, \AIST, \ASP, \COMMIT, \EVI), where do they live in the stack, and what concrete control actions do they trigger?

\begin{figure*}[t]
\centering
\includegraphics[width=1\textwidth]{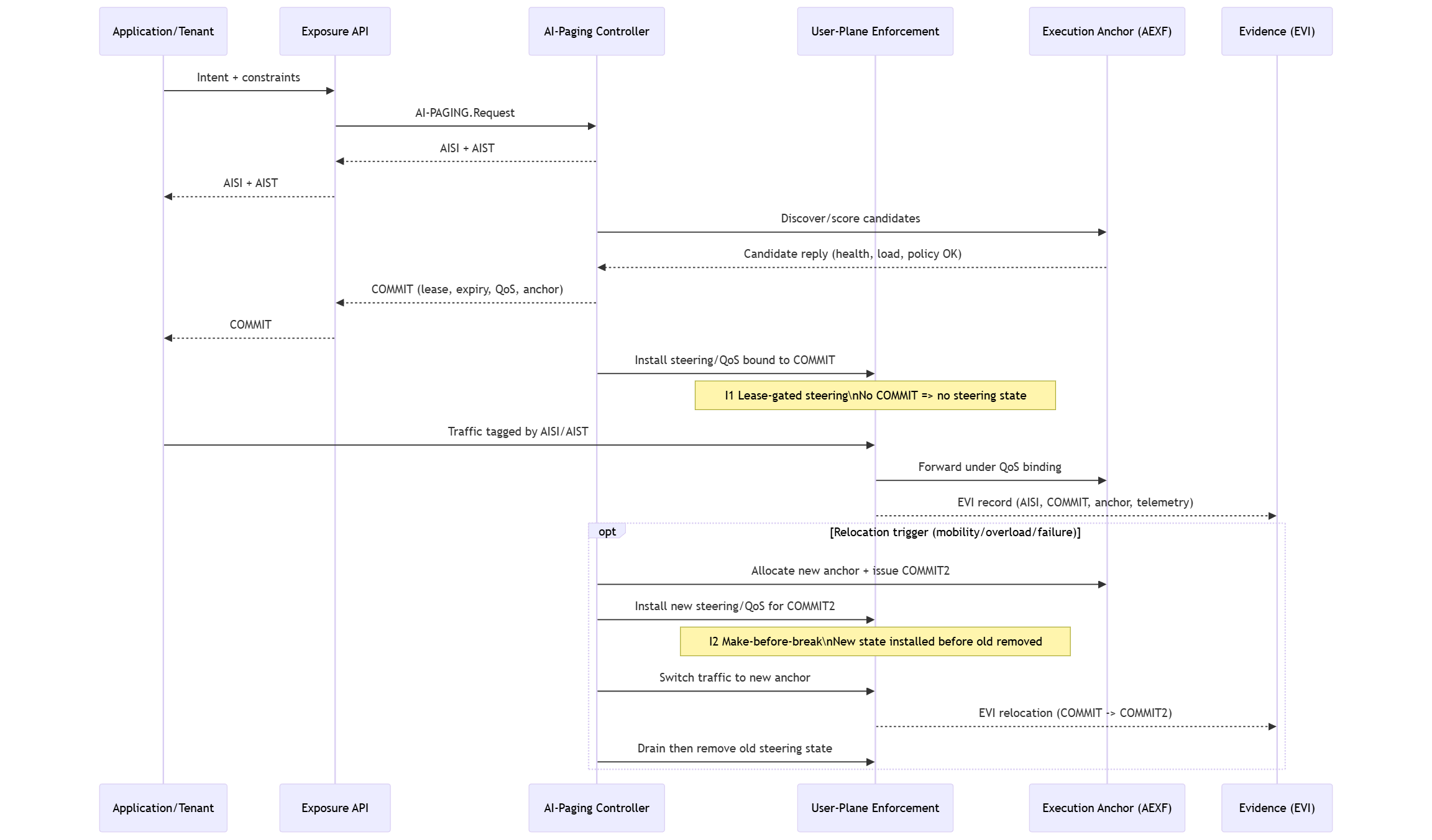}
\caption{System architecture mapping for AI-Paging. The exposure API terminates the intent and returns a stable handle (AISI/AIST). The AI-Paging controller, locating in the service layer (SEAL-like), adjacent to the exposure gateway (CAPIF/NEF-facing), or as part of the operator AIaaS orchestrator. derives ASP, selects candidates, and acquires a time-bounded COMMIT. The user plane installs steering and QoS state only when backed by an active COMMIT and forwards traffic to the admitted execution anchor. Telemetry and anchor events are bound into EVI records for audit-ready attribution.}
\label{fig:proto_map}
\end{figure*}

Fig.~\ref{fig:proto_map} depicts the prototype mapping from AI-Paging artifacts to enforceable control/user-plane mechanisms.
At a high level, \AIPaging\ is implemented as a control-plane service that (i) terminates the northbound intent call through an exposure framework (CAPIF-style onboarding/authz), (ii) orchestrates admission leases across an execution substrate (edge/cloud) and network delivery treatment, and (iii) programs steering and \QoS\ state through existing session/traffic-control hooks.
The user plane then forwards traffic according to lease-backed state, while the evidence pipeline binds delivery observables to the active lease and anchor.

\subsection{Control-plane realization: intent, identity, and admission leases}
The control-plane entry point is a network-exposed API call that carries the application intent and receives a stable service handle in return.
CAPIF provides the natural substrate for this interaction by standardizing API onboarding, discovery, and authorization for northbound interfaces~\cite{3gpp_capif_ts23222,etsi_ts123222}.
In the prototype, \AISI\ is realized as the stable handle returned to the invoker (e.g., a service URI or opaque identifier), while \AIST\ is realized as a scoped authorization token bound to \AISI\ and policy constraints (expiry, allowed tier/locality).
These choices are deliberately conventional: they reuse standard API security patterns so that the novelty stays in the lease semantics rather than in authentication machinery.

The admission lease \COMMIT\ is realized as a time-bounded authorization that ties together two otherwise independent decisions: (i) anchor-side capacity admission (compute feasibility for the chosen model tier), and (ii) network-side enforcement authorization (permission to install steering and \QoS\ state toward that anchor). NWDAF provides analytics not only for the evidence pipeline but also for candidate ranking and selection.
From the perspective of the control plane, the key is that \COMMIT\ is not merely a database record: it is treated as the only authority that enables enforcement state to exist, and its expiry is operationally meaningful (i.e., it triggers state removal). This is exactly what converts best-effort steering into enforceable service behavior.

\subsection{User-plane enforcement: steering and \QoS\ binding without new headers}
The prototype does not require a new packet header.
Instead, it assumes that traffic belonging to an \AIaaS\ service can be classified using a stable session-level classifier that is matchable by the enforcement point (e.g., by relying on the existing 5G application identifier and traffic detection mechanisms, allowing the UPF to classify traffic without introducing new headers, subject to accurate traffic detection and policy configuration).
The only requirement is deterministic mapping from user-plane traffic to (\AISI, \AIST) and thus to an active \COMMIT.

Once the lease is accepted, steering and \QoS\ binding are installed using existing programmable policy hooks.
In a 5G-oriented deployment, this corresponds to configuring user-plane forwarding and QoS-flow treatment under the 5GS architecture and procedures~\cite{3gpp_5gs_arch_23501,3gpp_5gs_proc_23502}.
At an abstract level, steering determines the selected execution anchor \AEXF\ (edge/cloud endpoint) and the forwarding path toward it, while \QoS\ binding determines the treatment class for the corresponding flow (e.g., latency-appropriate scheduling and policing parameters).
The crucial semantic is not the specific mechanism used by a given vendor, but the lease gate: installation is permitted only when a valid \COMMIT\ exists, and removal follows lease expiry or revocation deterministically.


\subsection{Operational bounding: timeouts, revocation, and make-before-break}
Two operational boundaries are enforced in the prototype.
First, admission is time-bounded by a commit timeout, which limits how long the controller attempts candidate anchors or permitted tier fallbacks before returning failure.
Second, relocation is bounded by a drain timeout, which controls make-before-break overlap: the new lease and enforcement state are installed and activated before the old path is drained and released.
These timeouts are not implementation details; they are part of the service semantics since they bound time-to-admit and worst-case relocation overlap.

Lease expiry and revocation are treated as first-class events.
When the active \COMMIT\ expires, enforcement state is removed; when the controller revokes a lease (e.g., policy change or detected abuse), steering and \QoS\ state are withdrawn accordingly.
This is what makes ``no lease, no steering'' testable: the prototype can log the mapping between lease validity and the presence/absence of enforcement rules.

\subsection{Evidence pipeline: binding delivery to lease state}
Evidence is produced by combining user-plane measurements and anchor-side events into \EVI records bound to \AISI\ and the active lease identifier.
On the network side, telemetry/analytics functions can supply delay/jitter/loss counters and flow-level observables; 5G analytics services such as NWDAF illustrate how such telemetry can be exposed to control functions~\cite{3gpp_nwdaf_ts23288}.
On the anchor side, the execution substrate contributes service-level signals such as queueing delay, model tier selected, and relocation events.
The design does not require disclosure of internal topology or proprietary schedulers; it requires attributable evidence tied to the service handle and lease state so that violations and transient degradations (e.g., during relocation) can be interpreted and audited.

This prototype mapping sets up the evaluation that follows.
since the control plane produces explicit artifacts and the user plane enforces lease-backed state, the evaluation can measure not only average performance but also correctness properties: whether steering ever exists without a valid lease, how quickly enforcement is withdrawn upon expiry, and how relocation affects service interruption under mobility, overload, and anchor failures.

\section{Evaluation}

The evaluation is designed to answer a simple question: does AI-Paging turn \AIaaS\ from ``endpoint selection + best-effort steering'' into an enforceable network service, without introducing prohibitive control overhead?
Rather than optimizing for a particular placement policy, the experiments focus on the semantics introduced by \AIPaging: (i) whether admission leases actually gate enforcement in the user plane, (ii) whether make-before-break relocation preserves continuity under realistic churn, and (iii) whether evidence generation is practical and useful for accountability.

\subsection{Experimental setup and baselines}
The prototype is deployed across a multi-anchor environment with at least one edge execution site and one remote execution site, connected through a programmable user-plane enforcement point capable of installing steering and \QoS\ state.
The control plane exposes the intent API, runs the AI-Paging controller, and implements lease issuance, expiry, and revocation.
Execution anchors implement anchor-side admission (capacity acceptance/rejection) and emit anchor-side events needed for evidence binding (queueing delay, model tier chosen, relocation events).

The evaluation compares against two baselines that represent common deployment choices.
The first baseline binds clients to a fixed endpoint selected at session start and uses application retries on failure (endpoint-bound serving).
The second baseline allows steering changes but does not gate installation on an admission lease (best-effort steering).
The proposed design is lease-backed steering with transactional relocation, where enforcement state is permitted only when a valid COMMIT exists.



\begin{figure}[t]
    \centering
    \includegraphics[width=0.98\linewidth]{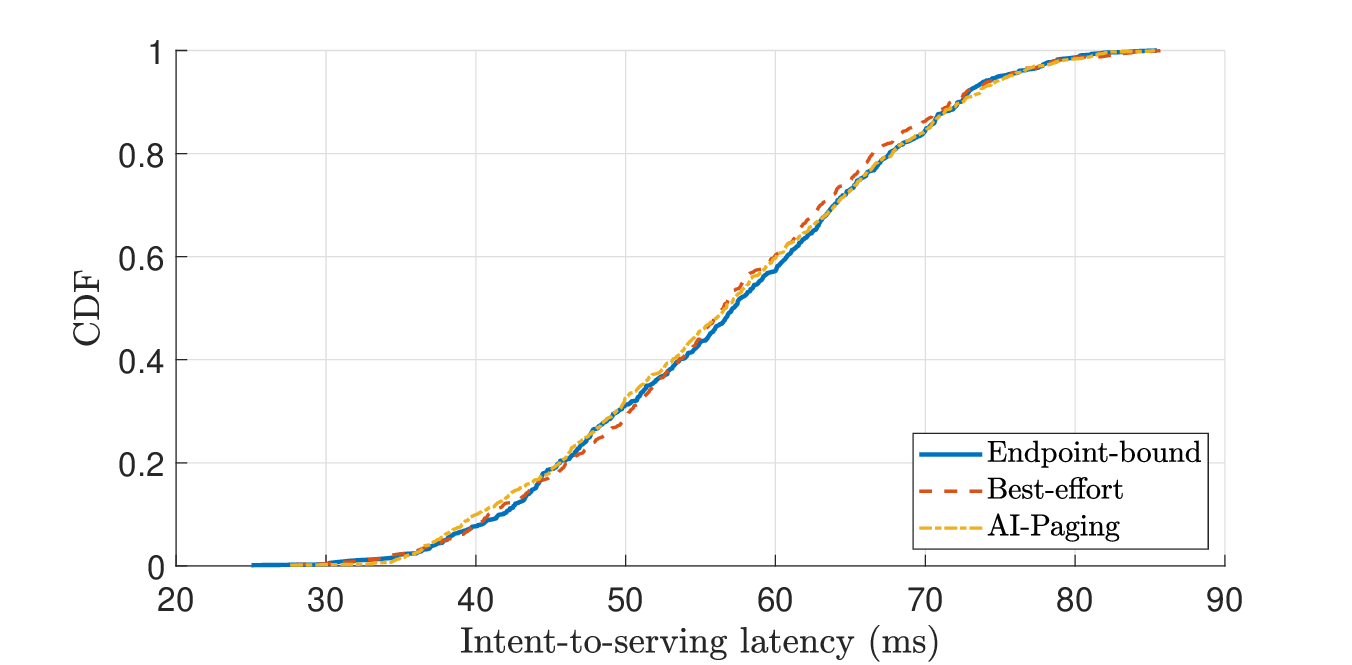}
    \caption{Intent-to-serving transaction time across designs.}
    \label{fig:txn_time}
    \vspace{-0.4cm}
\end{figure}

\begin{figure}[t]
    \centering
    \includegraphics[width=0.98\linewidth]{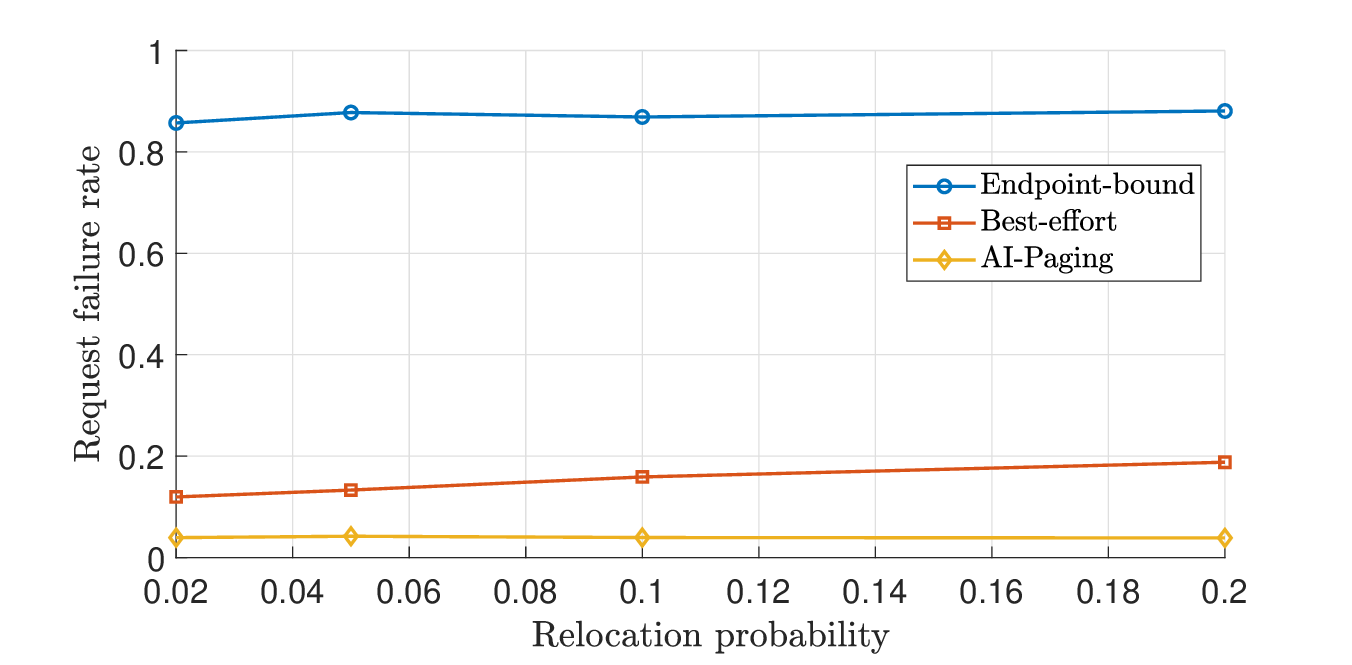}
    \caption{Relocation continuity metrics under mobility/churn conditions.}
    \label{fig:reloc_cont}
     \vspace{-0.4cm}
\end{figure}

\begin{figure}[t]
    \centering
    \includegraphics[width=0.98\linewidth]{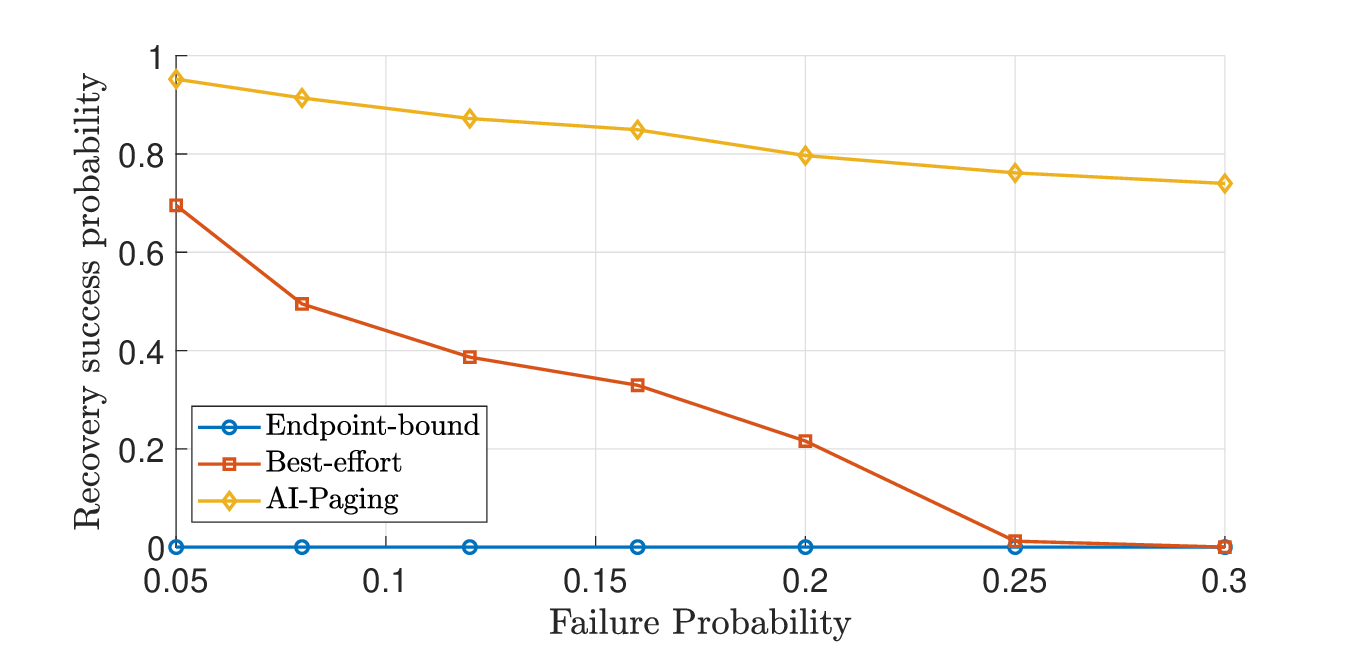}
    \caption{Recovery success probability versus stress level (offered load/churn).}
    \label{fig:recovery_prob}
    \vspace{-0.4cm}
\end{figure}

\begin{figure}[t]
    \centering
    \includegraphics[width=0.98\linewidth]{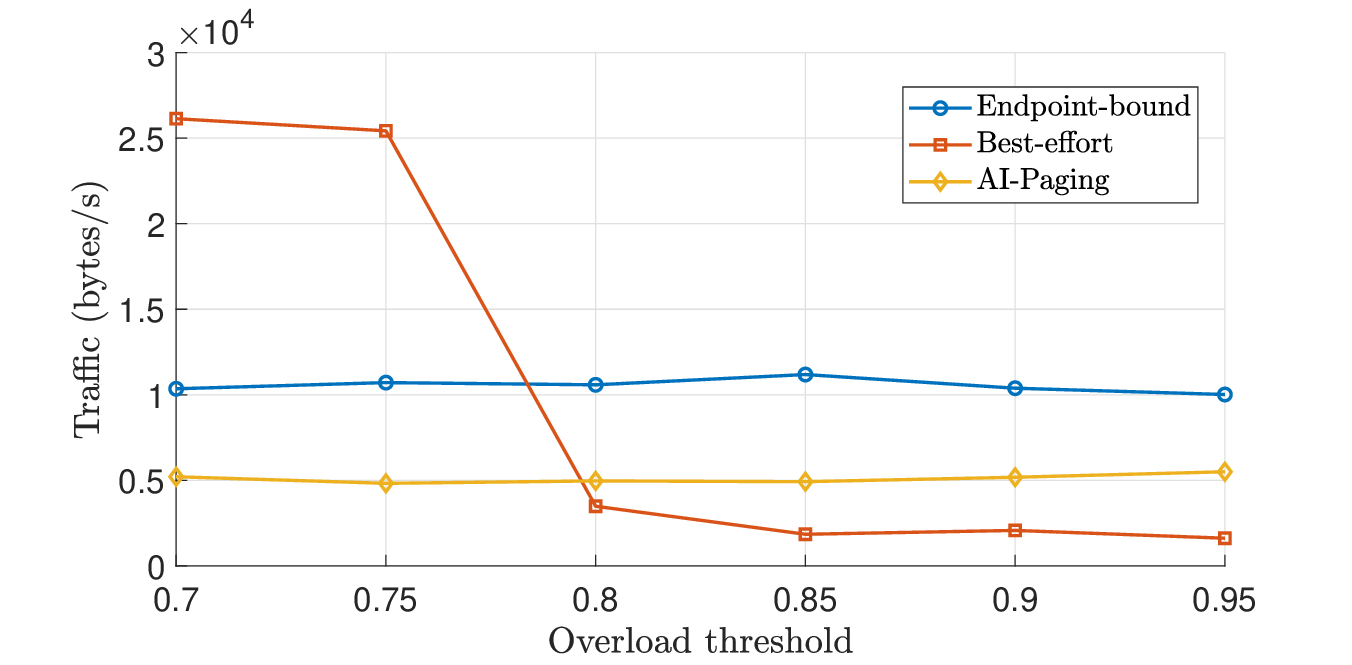}
    \caption{Evidence traffic rate versus overload threshold.}
    \label{fig:evi_traffic}
    \vspace{-0.4cm}
\end{figure}

\subsection{Results}

 Network dynamics are emulated by changing path conditions and reachability in a controlled manner, forcing the system to relocate between anchors while preserving the stable service handle. Overload is injected by reducing anchor admission capacity or increasing request arrival rate until the preferred anchor begins rejecting admission, exercising bounded fallback and permitted tier degradation. Failures are injected by removing an anchor (hard failure) or degrading its health signal (soft failure), measuring time to recover service via an alternate admitted lease. Lease expiry is exercised by issuing short leases and validating that enforcement state is removed deterministically on expiry, and that traffic is not silently steered once admission is no longer valid. Each scenario is executed repeatedly with randomized seeds for arrival patterns and failure timing.

\subsubsection{Distributional behavior under churn, overload, and failure}
Fig.~\ref{fig:txn_time} (intent-to-serving transaction time CDF) shows that AI-Paging remains in the same latency regime as EndpointBound and BestEffort, despite introducing explicit lease semantics. The three CDFs are closely aligned over most quantiles, which indicates that adding admission-backed enforceability does not create a large control-plane latency penalty. This is important: the proposed mechanism is not trading correctness for impractical setup overhead; instead, it preserves transaction-time practicality while adding deterministic control semantics.

Fig.~\ref{fig:reloc_cont} reveals stronger separation under relocation churn. The request-failure-rate curve for AI-Paging stays near zero across the relocation-probability sweep, while BestEffort increases in the low-to-moderate range and EndpointBound remains very high across the full range. This pattern directly supports the make-before-break interpretation: relocation succeeds because the target anchor is admitted and installed before old-path teardown, so continuity is preserved even as churn probability rises. In contrast, non-transactional or endpoint-bound strategies are more exposed to transient gaps, retries, and admission mismatches during handover episodes.

Fig.~\ref{fig:recovery_prob} provides the strongest robustness signal under compounded stress (offered load/churn). All methods degrade as stress increases, which is expected; however, the degradation profile is materially different. AI-Paging retains a high recovery-success level over the practical operating region and degrades gradually. BestEffort deteriorates much faster and approaches collapse at higher stress values. EndpointBound stays near the floor throughout, showing that static endpoint affinity is fundamentally fragile when failures and resource pressure become frequent. Operationally, this means AI-Paging does not merely improve average outcomes; it enlarges the usable stress envelope before saturation.

Fig.~\ref{fig:evi_traffic} clarifies evidence-plane overhead behavior. AI-Paging exhibits a controlled and relatively stable evidence traffic profile as the overload threshold varies, while BestEffort shows high sensitivity and a sharp regime change around intermediate thresholds. EndpointBound appears more stable but at a higher overhead level than AI-Paging in most of the sweep. The key implication is that lease-backed relocation and explicit state transitions generate evidence in a predictable way, which is preferable for auditability and capacity planning. In other words, accountability signals scale in a bounded and operationally tractable manner rather than becoming trigger-noise dominated.

Taken together, Figs.~\ref{fig:txn_time}--\ref{fig:evi_traffic} establish a coherent causal chain: AI-Paging adds transactional control semantics with minimal setup-time disruption (Fig.~\ref{fig:txn_time}), reduces continuity failures under relocation churn (Fig.~\ref{fig:reloc_cont}), preserves recovery capability under high stress (Fig.~\ref{fig:recovery_prob}), and keeps audit/evidence overhead controlled (Fig.~\ref{fig:evi_traffic}). This is precisely the behavior expected from lease-gated steering plus make-before-break relocation.

\subsubsection{Enforcement correctness and operational implications}

Table~\ref{tab:enforce_results_multi_setup} reports the enforcement-without-lease violation rate (percentage of simulated time where steering remains active without a valid lease). The separation is decisive across all setups. EndpointBound and BestEffort are non-zero in every case, ranging approximately from 17\% to 35\%, while AI-Paging is 0.000\% in all five scenarios. The average violation level of the two baselines is about 24.8\%, with the largest degradation under load-dominated setups (S3--S4), where values move into the mid-30\% range. By comparison, AI-Paging achieves a full elimination of the measured violation metric in the evaluated horizon.

\vspace{0.4cm}
\begin{table}[h!]
\centering
\footnotesize
\caption{Enforcement correctness across multiple setups.}
\label{tab:enforce_results_multi_setup}
\begin{tabular}{lrrrrrr}
\toprule
\textbf{Setup} & \textbf{EndpointBound} & \textbf{BestEffort} & \textbf{AIPaging}  \\
\midrule
S1 Nominal         & 18.127 & 18.093 & 0.000  \\
S2 High mobility   & 18.603 & 17.251 & 0.000  \\
S3 High load       & 35.263 & 34.933 & 0.000  \\
S4 Mobility+load   & 34.263 & 34.752 & 0.000  \\
S5 Failure stress  & 17.346 & 19.577 & 0.000 \\
\bottomrule
\end{tabular}
\vspace{0.4cm}
\end{table}

The table-level correctness evidence also explains the figure-level robustness trends. When steering state is strictly lease-gated, relocation and fallback decisions remain aligned with current admissibility, which reduces hidden mismatch between control-plane intent and user-plane behavior. That alignment is exactly what improves continuity and recovery under churn/failure stress and prevents pathological modes in which traffic continues to be steered toward anchors that are no longer valid for the requested service tier.

From a deployment perspective, the results support three practical conclusions. First, enforceability can be added without prohibitive transaction inflation (Fig.~\ref{fig:txn_time}). Second, continuity under realistic mobility/load/failure dynamics improves materially when relocation is transactional (Figs.~\ref{fig:reloc_cont} and \ref{fig:recovery_prob}). Third, evidence generation remains operationally manageable and policy-relevant (Fig.~\ref{fig:evi_traffic}) while enforcement correctness is preserved (Table~\ref{tab:enforce_results_multi_setup}). 

\section{Conclusion}
This paper argues that network-exposed \AIaaS\ requires carrier-grade service semantics, not just edge-hosted inference.
By introducing AI-Paging as a lease-based execution anchoring transaction, the design separates stable service identity from serving location, gates user-plane enforcement on explicit time-bounded admission, and preserves continuity through make-before-break relocation while producing evidence that binds delivery to the active lease and anchor.
The resulting contract is implementable using existing control/user-plane mechanisms and provides a reusable foundation on which optimization policies can improve without weakening enforceability. From a standardization perspective, AI-Paging could be specified as an AIMLE/CAMARA API behavior profile that defines normative lease semantics and enforcement expectations. Alternatively, it could be realized as a CAPIF-exposed service with standardized procedures governing lease creation, renewal, expiry, and relocation.

\bibliographystyle{IEEEtran}
\bibliography{refs}

\end{document}